\documentstyle[12pt]{article}

\begin{document}
\renewcommand{\arraystretch}{1.5}
\begin{center}
\begin{Large}
{\bf Effective Carrier Mean-Free Path in Confined Geometries}\\
\end{Large}
\bigskip
R. A. Richardson and Franco Nori\\
{\em Department of Physics, University of Michigan, Ann Arbor MI 48109-1120}
\bigskip
\bigskip
\begin{large}
{\bf Abstract}
\end{large}
\end{center}

\noindent
\baselineskip 4.9ex
The concept of exchange length is used to determine the effects of
boundary scattering on transport in samples of circular and rectangular
cross section.  Analytical expressions are presented for an effective
mean-free path for transport in the axial direction. The relationship
to the phonon thermal conductivity is discussed.

\newpage

The properties of transport in confined geometries has received
substantial attention in recent years (see {\em e.g.,} Ref.~[1]).  In
this Letter, the effects of boundary scattering on transport in small
samples is examined.   Our goal is to provide concise expressions for
the effect of boundaries on transport without explicitly evaluating the
Boltzmann equation. We present analytical expressions for the effective
mean-free path for axial transport in samples where the bulk mean-free
path is determined by other scatterers present in the sample.
Expressions are obtained for samples of circular and rectangular cross
section.  These results are applicable to samples which are small
enough that the carrier mean-free path is on the order of the sample
dimensions but not so small that the carrier spectrum is substantially
modified from the bulk.  In other words, the sample dimensions will be
assumed to be much greater than the carrier wavelength.

We employ a method first proposed by Flik and Tien \cite{flik} for the
calculation of the size effect in thin films. The method assumes that,
for a carrier of a given frequency in a bulk sample, a characteristic
mean-free path, $l$, can be defined.  The goal, then, is to examine how
this bulk value of $l$ is modified by the presence of boundaries in the
sample. The calculation utilizes the concept of the exchange length
$l_{ex}$ \cite{flik,tien}, which is defined as the average distance
normal to a plane that a carrier travels after having been scattered
within that plane. Specifically, we consider a carrier that has
undergone a scattering event within a plane that is perpendicular to
the direction of net transport which will be referred to from here on
as the positive $z$ direction.
We now allow the carrier to propagate to the point of its next
scattering event, which, in the bulk, is a distance $l$ away.  This
propagation is assumed to proceed with equal likelihood in all
directions. $l_{ex}$  is then defined as the average z-component of all
possible such propagation vectors, where the average is performed over
the hemisphere in the positive $z$ direction.  The bulk value of this
quantity, $l_{\infty}$ is determined by the expression \cite{flik},

\begin{equation}
l_{\infty} =  \frac{1}{2 \pi} \int_{0}^{2\pi}
d\phi \int_{0}^{\pi /2} d \theta \,l \sin\theta \, \cos\theta  = \frac{l}{2},
\end{equation}
where $\theta$ is the angle between the vector {\large \boldmath $l$}
and the $z$ axis and $\phi$ is the angle between the projection of
{\boldmath $l$} into the $xy$ plane and the positive $x$ axis.

In the following, we consider the scenario in which the mean-free path
is on the order of the sample dimensions.  For this case, some carriers
will strike the boundaries before traveling a full distance $l$ and the
exchange length will be correspondingly shorter. We will assume that
scattering at the boundaries is  diffuse, which will be valid when the
carrier wavelength is smaller than the characteristic roughness
features of the sample surface.  This hypothesis needs to be examined
within the context of a particular measurement, but holds for many
materials. We note that a principle purpose of this work is to extend
the most widely used form for the effects of boundary scattering on the
thermal conductivity \cite{casimir} and in this form, perfect sample
roughness is also assumed.
We will also consider the sample to be
free of grain boundaries, though the expressions derived could perhaps
also be applied to samples whose grains have characteristic geometries
which match those investigated here. A geometric analysis will be used
to calculate the average exchange length for axial transport in narrow
samples (length is assumed infinite) of both circular and rectangular
cross section.  This quantity will then be related to an effective
mean-free path, $l_{\em eff}$, which can be used to calculate axial
transport coefficients.

In order to calculate the exchange length for a cylindrical sample, we
initially assume that the excitation can originate with equal
likelihood anywhere within a given circular cross section of the
sample.  The average value of the exchange length in the sample, $
\tilde{l}_{ex} $, is then  obtained by averaging $l_{ex}$ (which is
itself an average over a hemisphere of solid angle) over the entire
cross section.
The geometry to be considered is shown in Fig.~1.  We consider an
excitation originating at some point a distance $\rho$ from the center
of the cross section of radius $R$ and propagating in a random
direction within the hemisphere of solid angle whose base is normal to
the positive $z$ direction. The quantity $\theta$ is defined as the
angle between the propagation vector, {\boldmath $l$}  and the z-axis,
and $\phi$ is the angle between the radius along which the origination
point  is located and the projection of  {\boldmath $l$}  into the
plane of the cross section. Note that  {\boldmath $l$}  may or may not
have length $l$, depending upon whether or not it is truncated by a
boundary.   The average exchange length is then  given by the
expression,

\begin{equation}
 \tilde{l}_{ex}  =\frac{2}{\pi R^2} \int_{0}^{R} d \rho   \rho
  \int_{0}^{\pi}d\phi
 \int_{0}^{\pi/2}d\theta \sin{\theta} \,l_z( \rho ,\phi,\theta),
\end{equation}
where $l_z$ is the $z$ component of the propagation vector heading in
the $(\theta,\phi)$ direction and $\phi$ has only been integrated over
half its range for symmetry reasons.  In the above, an integration over
the polar angle $\xi$ within the cross section plane has already been
performed since $l_z$ does not depend on it.

The evaluation of this integral involves a careful analysis of the
regions of $(\rho,\phi,\theta)$ space where {\boldmath $l$} does and
does not hit the wall.  This is of importance since the functional form
of $l_z$ clearly depends on whether or not  {\boldmath $l$} is
truncated by a collision with a wall. The details of this process are
beyond the scope of this article and will be presented in a more
comprehensive work \cite{me}. For the present discussion, suffice it to
say that one needs to consider two distinct regimes: $l \geq 2R$ and $l
< 2R$.  The expressions for $\tilde{l}_{ex}$ for these two cases are
given in Table I.

We have also evaluated the average exchange length for samples of
rectangular cross section.  The general expression for $\tilde{l}_{ex}$
is quite analogous to the circular case and is given by

\begin{equation}
 \tilde{l}_{ex}  =\frac{2}{\pi ab} \int_{0}^{a/2}
 dx \int_{0}^{b/2} dy  \int_{0}^{2\pi}d\phi
 \int_{0}^{\pi/2}d\theta \sin{\theta} \,l_z(x,y ,\phi,\theta),
\end{equation}
where $a$ and $b$ are the lengths of the two sides and the integration
is performed over only the bottom-left quarter of the cross section for
symmetry reasons.  The  bottom-left corner is chosen as the origin of
$x$ and $y$ and the $a$ side lies along the $x$ direction.

As in the circular case, the  evaluation of Eq.~(3) depends upon
discerning which regions of phase space have  {\boldmath $l$} hitting
the wall and which do not.  The analysis of this problem will be
presented in detail in Ref.~\cite{me}.  Because of the reduced symmetry
of this case relative to the circular one, the solution needs to be
broken down into more regimes. Specifically, we need to consider the
cases where $l$ is: less than both dimensions, greater than the shorter
side but less than the longer, greater than both sides but less than
the diagonal, and, finally, greater than the diagonal.  All of these
results are shown in Table I.

The expressions presented thus far for the evaluation of
$\tilde{l}_{ex}$, for both the circular and rectangular cases, were
derived on the assumption of uniform origination, {\em i.e.,} we assume
that the excitation can originate anywhere within the cross section
with equal likelihood.  Recalling the physical interpretation of the
exchange length, however, we note that the excitations we are
considering are those that have just undergone a scattering event
within the cross section plane.  As pointed out in Ref. \cite{flik},
when the mean-free path of the excitation becomes much longer than the
sample dimensions, the excitation becomes increasingly likely to
scatter on a boundary and our assumption of uniform origination needs
to be replaced with a boundary origination description.
The calculation of the boundary origination solutions is
straightforward and the expressions for the two geometries are as
follows: Circular:  $\tilde{l}_{ex}  = \frac{4R}{\pi} - \frac{R^2}{l}$;
Rectangular:  $\tilde{l}_{ex} = J$ where $J$ is given in Table I.

The appropriate procedure \cite{flik} is to match these solutions to
those for uniform origination at large $l$'s.   This can be
accomplished by a simple exponential matching process, whereby the
matched solution is obtained by adding the uniform origination solution
to the difference between the boundary and uniform solutions times an
exponential function.  This process results in the equations in Table I
where $l\geq 2R$ for the circular case and $l \geq d$ for the
rectangular case.

These forms result in small discontinuities when they are combined with
the solutions for $ l < 2R$  and $l < d$, respectively.  The matched
solutions contain a matching parameter in the exponential functions
which we choose to be four to achieve the best compromise between
minimizing this discontinuity and ``phasing in" the boundary
origination solution as quickly as possible as $l$ increases. It should
be noted that, for the rectangular case when $a \gg b$,  we can expect
to be largely in the boundary origination regime before $l$ exceeds the
length of the diagonal.  The above equation does not allow for this
possibility and is therefore most applicable to cases in which $a$ is
not too different from $b$.  For samples where one dimension is much
larger than the other, we refer the reader to the results of Ref.
\cite{flik} where an expression for thin films is derived.

The expressions given in Table I can now be used to aid in the
calculation of transport quantities in samples where boundary
scattering is expected to play a role.  The general procedure is as
follows.  We know that the bulk value of the exchange length is given
by $l/2$. In other words, the bulk mean-free path is given by two times
the exchange length. Now, this relationship can be extended to the
confined geometry case by defining an effective mean-free path for
axial transport as $l_{\em eff} = 2\tilde{l}_{ex}$, where
$\tilde{l}_{ex}$ for axial transport is given by the expressions in
Table I.

Let us apply this relationship to the thermal conductivity.  In the
kinetic theory approximation, the thermal conductivity $\kappa$ is
given by $\kappa = \frac{1}{3} C v l$, where $C$ is the contribution to
the specific heat from the carrier in question and $v$ is the carrier
velocity.  Now, in Ref. \cite{flik}, it is asserted that, within this
approximation, the transport $\kappa_z$ along the $z$ axis in a sample
of confined geometry can be obtained from the bulk thermal conductivity
$\kappa_{\infty}$ and the exchange length along that axis by the
expression $\kappa_z = \kappa_{\infty} \frac{\tilde{l}_{ex}}{l/2}$.
It can easily be seen then that the value of the thermal conductivity
for the confined sample is obtained from the bulk expression by simply
substituting $l_{\em eff}$ for $l$.  In other words, for a small sample
with a bulk mean-free path $l$, $\kappa_z(l) = \kappa_{\infty}(l_{\em
eff})$, where $l_{\em eff}$ is defined above.

Because of the simple geometric nature of this argument, it is
plausible that this sort of analysis can be applied to more
sophisticated treatments of the thermal conductivity as well.  For
instance, in the case of phonon transport,  the thermal conductivity is
often written in the Debye approximation as an integral over phonon
frequencies \cite{berman};
\begin{equation}
\kappa_p(T)={k_B\over 2\pi^2v}({k_B\over \hbar})^3 T^3
\int_0^{\theta_D/T} dx\,{{x^4\,e^x}\over (e^x-1)^2}\, \tau(T,x),
\end{equation}
where $x$ is the reduced phonon frequency $\hbar \omega /k_B T$, $k_B$
is the Boltzmann constant, $\theta_D$ is the Debye temperature and
$\tau(T,x)$ is the frequency dependent scattering time.  The total
inverse scattering time,  $\tau(T,x)^{-1}$, is usually expressed as a
summation of the inverse scattering times from scatterers of various
types, {\em e.g.}, point defects, phonon-phonon umklapp processes,
etc.  Within this context, the effect of boundaries is typically
handled by a method due to Casimir\cite{casimir} whereby one adds a
frequency independent term to this total of the form  $\tau_{b}^{-1} =
v/\alpha d$, where $d$ represents the sample dimension and $\alpha$ is
a geometrical factor.

We propose that greater accuracy may be achieved from calculations
involving Eq.~(4) by omitting the boundary scattering term in the total
inverse scattering time and utilizing the  equations in Table I for the
exchange length to modify the mean-free path instead.  The proposed
procedure is as follows:  the factor $\tau$ in Eq.~(4) can readily be
replaced by $l(x,T)/v$, at which point the integration can be seen to
be over the frequency dependent mean-free path multiplied by another
$x$ dependent factor.  For each such mean-free path, $l(x,T)$, a value
of $\tilde{l}_{ex}$ can be derived by using the equation appropriate
for the geometry of the particular sample under investigation.  Each
$l(x,T)$ in the integral can then be replaced by an $l_{\em eff}(x,T) =
2\tilde{l}_{ex}(x,T)$ as described above.  The proper, boundary limited
value of the thermal conductivity is then obtained by integrating over
the $l_{\em eff}(x,T)$, with the other factors in Eq.~(4) left
unaltered.

An example of this process is pictured in Fig.~2.  The graph shows two
calculations of the thermal conductivity of diamond using Eq. 4 as a
basis.  The solid curve is taken from a recent work\cite{onn} and uses
the simple treatment whereby the boundary scattering is handled by the
addition of a constant term to the inverse scattering time
\cite{casimir}.  The form of the constant term is for axial transport
along perfectly rough grains of square cross section ($\alpha =
1.12$).  The model also includes a point defect scattering term and a
phonon-phonon umklapp term.  The dashed curve shows the results when
all of the parameters of the model are left unchanged but the boundary
scattering is treated by the method described in this work (the square
cross section equations are used).
The assumptions about direction of transport and surface roughness are
the same as for the solid curve.  One can see that the present method
produces a significant enhancement of the thermal conductivity peak
relative to the Casimir method.

In summary, we have presented analytical expressions for the effects of
boundary scattering in samples where the bulk carrier mean-free path is
determined by other scatterers present.  Results are derived for axial
transport in long, narrow samples of circular and rectangular cross
section, where scattering at the boundaries is diffuse.  The results
are incorporated into a definition of an effective mean-free path for
axial transport which can be used to calculate coefficients such as the
thermal conductivity.  Though we have focused on thermal transport in
the present work, the expressions derived here could be of use in the
examination of a variety of transport phenomena in confined
geometries.

\medskip
FN acknowledges partial support from a GE fellowship, a Rackham grant,
the NSF through grant DMR-90-01502, and SUN Microsystems.

\newpage
\noindent
\begin{large}
{\bf Figure Captions}
\end{large}

\noindent
Figure 1:  Schematic diagram showing the geometry relevant for the
calculation of the exchange length for a wire of circular cross
section.  The case pictured is that for which {\boldmath $l$} hits the
boundary.

\bigskip
\noindent
Figure 2:  Calculation of the thermal conductivity of diamond using the
current and Casimir methods.  The present technique produces a 12$\%$
enhancement in the peak relative to the Casimir method.  Model
parameters are taken from Ref. \cite{onn}.


\newpage
\begin{thebibliography}{99}

\bibitem{rossi} F.~Rossi, L.~Rota, C.~Bungaro,
P.~Lugli, and E.~Molinari, Phys. Rev. B  {\bf 47}, 1695
(1993); V. B. Campos, S. Das Sarma, and M. A. Stroscio, {\em ibid}
{\bf 46}, 3849 (1992); S.--F. Ren and Y.--C. Chang, {\em ibid}   {\bf
43}, 11857 (1991); M. A. Stroscio, {\em ibid}  {\bf 40}, 6428 (1989);
J. Seyler and M. N. Wybourne, Phys Rev. Lett. {\bf 69}, 1427 (1992).

\bibitem{flik} M. I. Flik and C. L. Tien, Journal of Heat Transfer
{\bf 112}, 872 (1990).

\bibitem{tien}  C. L. Tien and J. H. Leinhard,
 {\em Statistical Thermodynamics},  pp. 307-321 (Hemisphere, New York, 1979).

\bibitem{casimir} H. B. Casimir, Physica {\bf 5}, 495 (1938).

\bibitem{me} R. A. Richardson and F. Nori, unpublished.

\bibitem{berman} See for example, R. Berman,
{\em Thermal Conduction in Solids} (Clarendon Press, Oxford, 1976).

\bibitem{onn} D.~G.~Onn, A.~Witek, Y.~Z.~Qiu, T.~R.~Anthony, and
 W.~F.~ Banholzer, Phys. Rev. Lett. {\bf 68}, 2806 (1992).

\end{thebibliography}
\end{document}